\documentclass[preprint,review,12pt]{elsarticle}

\usepackage{graphicx}
\usepackage{amssymb}
\usepackage{amsthm, amsmath}
\usepackage{xcolor}

\journal{X}

\begin{document}

\begin{frontmatter}


\title{Precise characterization of polarizing prisms by optical localization analysis}



\author[1]{Hoi Chun Chiu} 
\author[2]{Zhuohui Zeng} 
\author[3]{Luwei Zhao}
\author[3]{Teng Zhao}
\author[1]{Shengwang Du\corref{cor1}}
\author[2]{Xian Chen\corref{cor1}}
\cortext[cor1]{Corresponding authors: Xian Chen (xianchen@ust.hk) and Shengwang Du (dusw@ust.hk)}

\address[1]{Department of Physics, The Hong Kong University of Science and Technology, Clear Water Bay, Kowloon, Hong Kong, China}
\address[2]{Department of Mechanical and Aerospace Engineering, The Hong Kong University of Science and Technology, Clear Water Bay, Kowloon, Hong Kong, China}
\address[3]{Light Innovation Technology Ltd., Hong Kong, China}

\begin{abstract}
 We utilize the localization analysis method to precisely determine the light beam positions with the spatial separation beyond the optical diffraction limit. By such a direct spatial measurement, the associated optical setup is built to characterize the beam-shear angle of polarizing prisms that refract the light propagation according to polarization. We use the Nomarski prisms to demonstrate our method. The statistical error is below 0.5 $\mu$rad, which is two orders of magnitudes smaller than the tolerance value given by most prism manufacturers. Our method provides a stand-alone characterization of prisms with a high accuracy for many quantitative imaging technologies such as the orientation-free DIC microscope and the dual-focus fluorescence correlation spectroscopy. 
\end{abstract}


\end{frontmatter}


\section{Introduction}
In many light imaging systems, transparent prisms are widely used for refracting light. By means of applications, there are different types of prisms for deflecting or reflecting light beams. Among them, the polarizing prism which splits light into components with different polarizations, such as the Nomarski prism as a modified Wollaston prism \cite{Nomarski}, is the key optical component to generate the phase contrast for the differential interference contrast (DIC) microscope \cite{rosenberger1977differential}. Usually, the DIC microscope is exploited by researchers in material science and biology to observe the surface microstructure for opaque materials and the unlabeled structures for transparent materials. Many recent advances in the development of light imaging systems have utilized the optical beam-shear characteristic of the prism to quantify the surface roughness \cite{hartman1980quantitative}, and the microtopography of unlabeled living cells \cite{NguyenNC2017}. In some cases, it is not only used for imaging, but also serves as a functional module for the dual-focus fluorescence correlation spectroscopy to characterize the diffusion of molecules at nanomolar concentrations \cite{dertinger2007two, muller2008precise}. Beyond observation, these techniques enable the quantitative determination of the structural parameters and the associated physical properties, which has broadened the applications of the polarizing prisms towards multidisciplinary areas by a large margin. 

The optical beam-shear angle, as an important attribute of the polarizing prism, plays a crucial role in the quantitative measurement for all the aforementioned advanced techniques. However, neither the prism manufacturers nor the commercial DIC microscope companies provide the accurate specification of their polarizing prisms. It is common that the actual beam-shear angle of a Nomarski prism is approximately 20 $\mu$rad off to the nominal beam-shear angle tagged by the manufacturer. As a part of the imaging module, the specification of the Nomarski prism mostly goes with the specific objective lens, by which the beam-shear angle may be derived if the amount of lateral shear displacement at the focal plane is known. In this case, the measurement errors become uncertain, sometime are out of control when changing the objective lens or slightly modifying the illumination path of the microscope. 

Precise characterization of the beam-shear angle of a Nomarski prism is difficult because the amount of shear between the spatially separated light beams is usually quite small on the focal plane. The order of magnitude of the beam-shear angle of a typical Nomarski prism is from 10 to 100 $\mu$rad. It means that the corresponding shear displacement is of submicron scale on the focal plane. For most DIC microscopes, this value approaches the resolution limit of the light, which makes the direct spatial measurement impossible. Instead, several indirect methods are developed to estimate the beam-shear displacement from the phase information. One of the most straightforward method was demonstrated by Duncan {\it et al} when they was trying to quantify the light scattering and absorption of thin tissue cells \cite{DuncanJOSAA2011}. They used a wedge prism as the reference to calibrate the beam-shear displacement of the prism for their DIC microscope. The accuracy of the method depends on the orientation of the placed wedge, the precision of the wedge dimensions, and the coherence of the illumination source of their microscope. This approach is not universal because the measurement requires the whole microscope setting, and the measured data is sensitive to the coherence of the light source and fabrication errors of the wedge. The same problem arises for the similar methods using the dual-focus fluorescence correlation spectroscopy \cite{MullerOE2008}.

Some sample-less methods have been developed based on the interference between the light beams with small spatial separation. For highly overlapped light beams, depending on their coherence, the interference fringe pattern forms with certain periodicity. The amount of beam separation is related to the fringe spacing and wavelength. Such a relationship gives rise to determine the beam separation from the characteristics of the measured interference fringes. The accuracy of this method depends on the beam size and the intensity profile of the periodic fringes. The whole microscope setup is needed and the measurement errors are closely related to the specifications of objective lens, and the coherence of the laser source. Another reported interferometry approach utilizes the phase retardance between the sheared beams to quantify the amount of shear displacement at a preset shear plane \cite{ShribakJBO2008}. The precision of the measurement depends on the determination of the retardance and the position control of the prism along the transverse direction. 

For both sample-wise and sample-less methods, the errors arise from multiple sources including the optical setup of the microscope and the moving parts used in the measurement. By far, the quantitatively error analysis for various methods has not been reported. Nor has the bound of precision been derived. 
In this work, we proposed a method that directly determines the spatial positions for the light beams subjected to small lateral shear, by which we designed a stand-alone optical experiment to measure the beam-shear angle of polarizing prisms independent of the full microscope setup. During the data collection, there is no moving parts in our system. By optical localization analysis, it is possible to measure the spatial position for each of the light beams respectively. The only source of error comes from the stability of laser. We demonstrated our method using various Nomarski prisms and performed a statistical analysis for hundreds of independent measurements to determine the bound of precision. Finally, we formulated the theoretical precision for our method, on which a more accurate device can be built. 

\section{Principle and Optical Setup}

When two light beams of finite sizes are spatially close, it is difficult to resolve them due to the diffraction effect. Consider a collimated monochromatic light ({\it e.g.} laser) beam of wavelength $\lambda$, whose intensity profile is a Gaussian function, by which we define the beam diameter $D$ as the width of the Gaussian profile at the intensity of $1/{e^2}$.  After it passes through a polarizing prism, two orthogonally polarized light beams are produced and propagate along directions separated by a small shear angle $\varepsilon$. Restricted by the diffraction effect of light, such an angle has to be equal or larger than the diffraction angle of the beam, {\it i.e.},  $\varepsilon \geq \frac{4\lambda}{\pi D}$, so that the spatially sheared beams can be well resolved. This condition suggests that the beam size of incident light has to be sufficiently large ($D\geq\frac{4\lambda}{\pi\varepsilon}$) in direct observation to determine the optical beam shear angle. This picture can also be interpreted in the $\vec{k}$ space ({\it i.e.} the momentum space). The reciprocal distribution of the laser beam in the the $\vec{k}$ space has the width of $8/D$. For the two spatially sheared beams, their separation displacement in $\vec{k}$ space is $k \varepsilon$, where $k = \frac{2\pi}{\lambda}$ is the propagation number. The condition to resolve these two modes in the $\vec{k}$ space requires $k\varepsilon\geq8/D$, that is equivalent to the condition in real space $D\geq\frac{4\lambda}{\pi\varepsilon}$.  For example, in a typical configuration having the incident light of wavelength $\lambda = 400$ nm, it is required to have the beam size $D\geq5$ cm to resolve an optical beam shear angle $\varepsilon = 10$ $\mu$rad. It is unrealistic to utilize such a big beam size in experiment to measure the spatial variables directly since the diameter of most optical components is much smaller than that. 

To solve this problem, we apply the optical localization analysis \cite{STORM} to precisely measure the central positions of the two orthogonally polarized beams in their Fourier $\vec{k}$ space. In this way, it is possible to resolve the highly interfered light beams even below their diffraction limit because the two light beams do not overlap at all in the $\vec{k}-\hat{P}$ joint space. The localization method with distinguishable temporal modes has been used for the super-resolution fluorescence optical microscopy, such as photo-activated localization microscope \cite{PALM} and stochastic optical reconstruction microscope \cite{STORM}. Analogous to the principle of these super-resolution microscopes, we localize the position of an individual beam with respect to polarization so that the diffraction effect can be completely eliminated despite of the spatial closeness of the two light beams. 

\begin{figure}[ht]
\centering
\includegraphics[width=3.2in]{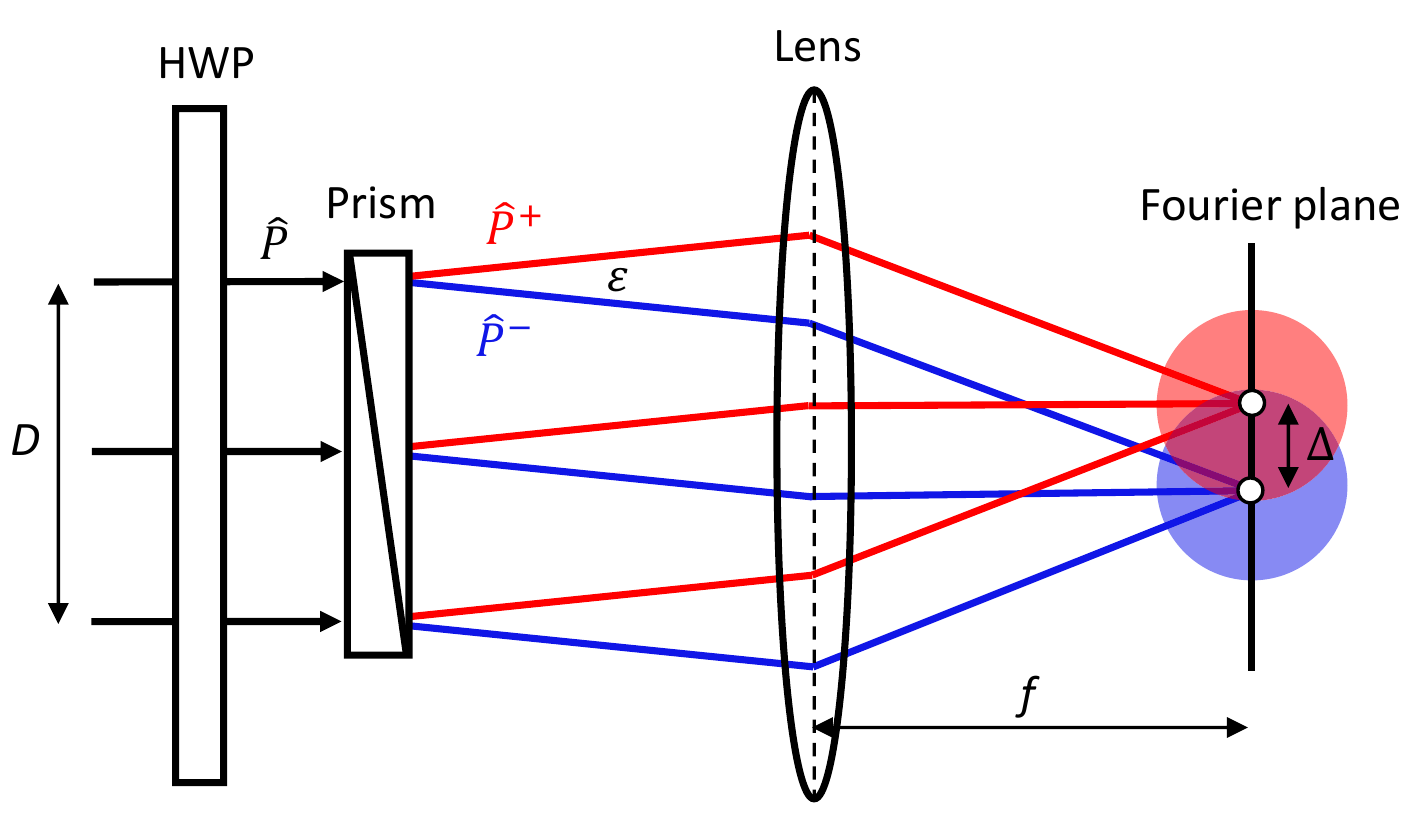}
\caption{Schematic of the optical setup for measuring the shear angle of a Nomarski prism. A linearly polarized and collimated laser Gaussian beam with a diameter $D$ is incident on the Nomarski prism. After passing the prism, two orthogonally polarized ($\hat{P}^+$ and $\hat{P}^-$) components are split by a shear angle $\varepsilon$, and focused by a lens on its front focal plane. The half-wave plate is used to localize each of the polarization states.} \label{fig1_os}
\end{figure}

We use the Nomarski prism as an example to illustrate the working principle by our measurement platform. The optical setup for measuring the beam-shear angle consists of a half-wave plate and a lens with focal length $f$ shown in figure \ref{fig1_os}. A linearly polarized and collimated laser beam ($\lambda=647$ nm, $D=3.56$ mm) passes though the half-wave plate and incidents into the Nomarski prism. The polarization of light before entering the prism is
\begin{equation}
    \hat P = \cos \alpha \hat P^+ + \sin \alpha \hat P^-, 
\end{equation}
where $P^+$ and $P^-$ denote two orthogonal polarization vectors. After the light passing through the prism, the light beam with $\cos \alpha \hat{P}^+$ is relatively sheared from the beam with $\sin \alpha \hat{P}^-$ by an angle $\varepsilon$. We used the half-wave plate to tune the value $\alpha$ of the incident beam so that the magnitude of either $\hat P^+$ or $\hat P^-$ will vanish in each of the independent data acquisition steps. The outgoing light beams from the prism ({\it i.e.} the red beam with $\hat P^+$ and the blue beam with $\hat P^-$) are focused by a lens with a focal length $f=15$ cm onto its front focal plane, which is equivalent to be viewed as the Fourier transform of the light to the momentum space. 

The distribution of localized intensity profile carrying a single mode of polarization collected at the Fourier plane is Gaussian, which is collected by the Nickon DS-Fi3 CCD camera with pixel size $2.4$ $\mu$m $\times$ $2.4$ $\mu$m. The shear displacement $\Delta$ (marked in figure \ref{fig1_os}), defined as the spatial distance between the central positions of the localized intensity profiles, is calculated as  
\begin{equation}
\Delta=f\varepsilon
\label{eq:Separation}
\end{equation}
where $f$ is the focal length of the lens and $\varepsilon$ is the beam-shear angle that is going to be measured by the proposed optical platform.
The size ($1/e^2$ diameter) of the intensity profile on the Fourier plane is calculated as
\begin{equation}
d=\frac{4\lambda f}{\pi D}.
\label{eq:SpotSize}
\end{equation}
If the two spots are imaged simultaneously, the criterion to resolve them is $\Delta\geq d$, which is limited by the diffraction effect. In order to avoid diffraction between the two beams, we tune the orientation of half-wave plate so that the polarization state of the incident beam is in a singular mode at different temporal steps. At $\alpha=0$, only the light polarized along $\hat{P}^+$ is present. We acquired the corresponding image showing the intensity $I^+(x,y)$ in figure \ref{fig:LocalizationMethod}(b). At $\alpha=\pi/2$, only the light polarized along $\hat{P}^-$ is present, of which the intensity is $I^-(x,y)$ shown in figure \ref{fig:LocalizationMethod}(c). Note that both $I^+$ and $I^-$ are Gaussian but their superposition is not, which is shown in figure \ref{fig:LocalizationMethod}(a). 
Compared with the intensity profile of the simultaneous spatial measurement of the beams on the focal plane, the independent intensity profile consisting of either $I^+(x,y)$ or $I^-(x,y)$ makes it possible to resolve the central positions of the highly overlapped intensity spots. 

\begin{figure}[ht]
    \centering
    \includegraphics[width=3.in]{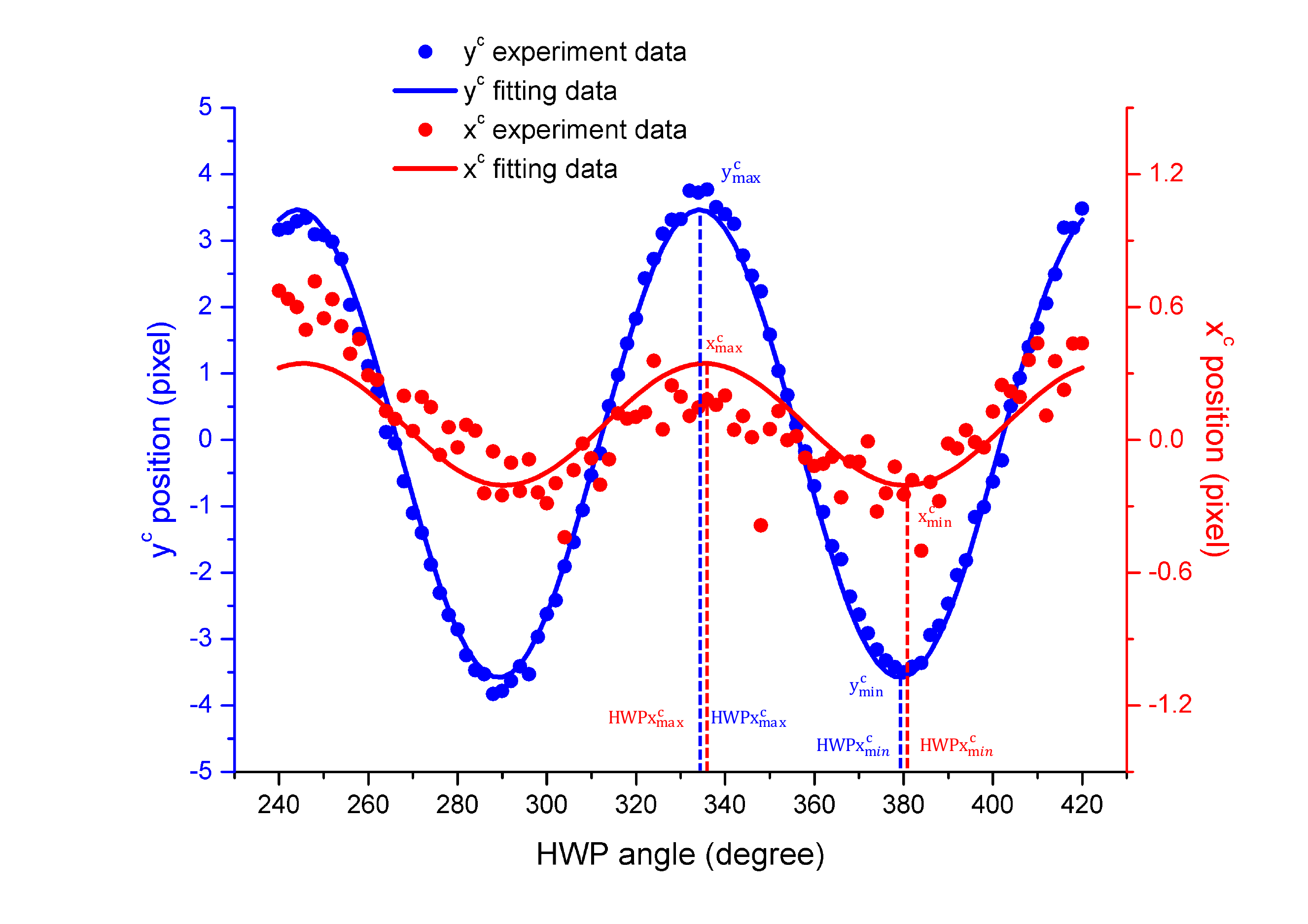}
    \caption{Variation of the center of mass of the intensity profile when tuning the angle of the half-wave plate.}
    \label{fig:cm}
\end{figure}

The angular alignment of the half-wave plate plays an important role in the determination of the spatial position of individual light beams. When the angle of the half-wave plate is set to be  $\omega$, the intensity profile after passing through the prism can be considered as the superposition of two Gaussian beams, {\it i.e. } $I(x, y; \omega)$. We define the center of mass of the intensity at $\omega$ as 
\begin{equation}\label{eq:cm}
    (x_\omega^c, y_\omega^c) = I_\omega^\ast (\int_{\mathbb R^2} x I dx dy, \int_{\mathbb R^2} y I dx dy)
\end{equation}
where $I_\omega^\ast = (\int_{\mathbb R^2} I dx dy)^{-1}$. As tuning the angle of the half-wave plate, the center of mass in \eqref{eq:cm} varies periodically between $(x_\text{min}^c, y_\text{min}^c)$ and $(x_\text{max}^c, y_\text{max}^c)$ as illustrated in figure \ref{fig:cm}. The finest angular scale of our half-wave plate is 2$^\circ$ ($\sim 0.035$rad). We chose the angles $\omega_1 = 335^\circ$ and $\omega_2 = 380^\circ$ for each of the data acquisition steps during the optical localization analysis. Physically,  rotating the half-wave plate by $|\omega_2 - \omega_1|$ causes the polarization of the incident light switching completely from $P^+$ to $P^-$.

The intensity profiles $I^+(x,y)$ and $I^-(x,y)$ corresponds to the outgoing beams with polarization $P^+$ and $P^-$ after passing through the polarizing prism. They are fitted by the 2-dimensional Gaussian functions
\begin{equation}
g^\pm(x,y)=I_0^\pm \exp\big[{-\frac{(x-x^{\pm})^2+(y-y^{\pm})^2}{2(\sigma^{\pm})^2}}\big]+I_b^\pm,
\label{eq:GaussianFunction}
\end{equation}
where the superscript $\pm$ corresponds to the state of polarization of the light, and $I_0^\pm$, $(x^\pm, y^\pm)$, $\sigma^\pm$ and $I_b^\pm$ are the fitting parameters. The spatial localization with respect to polarization allows for the determination of the central positions of the two modes in the $\vec{k}$ - $\hat P$ joint space. The central position of each of the intensity spots is represented by the calculated positions $(x^+, y^+)$ and $(x^-, y^-)$ from equation \eqref{eq:GaussianFunction}. The optical shear displacement of the two modes is determined as $\Delta=\sqrt{\Delta_x^2+\Delta_y^2}$ where $\Delta_x=|x^+-x^-|$ and $\Delta_y=|y^+-y^-|$. By equation (\ref{eq:Separation}), the beam-shear angle is given by
\begin{equation}
\varepsilon=\frac{f}{\Delta}.
\label{eq:ShearAngle}
\end{equation}

\begin{figure*}[ht]
\centering
\includegraphics[width=0.7\textwidth]{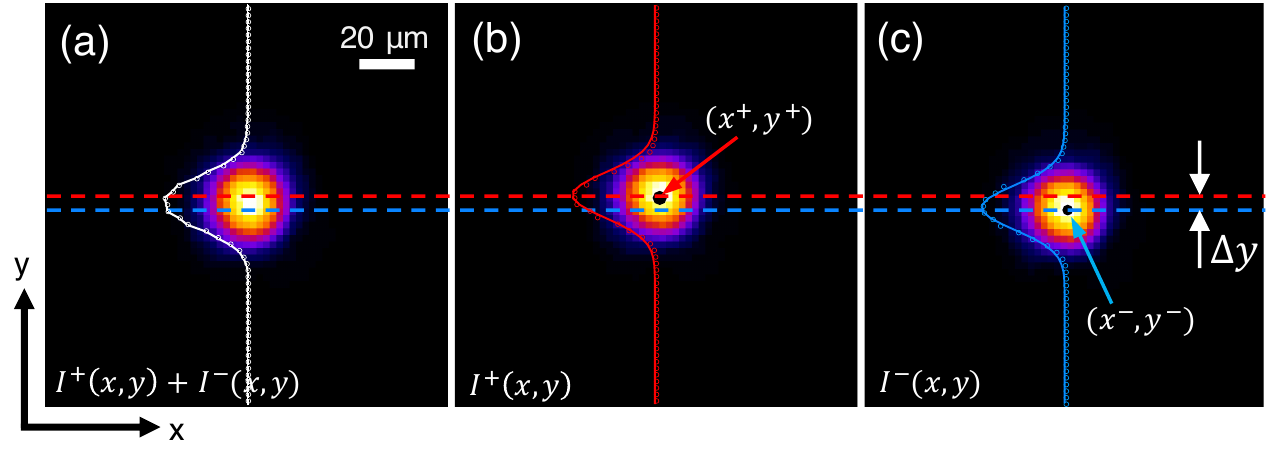}
\caption{Intensity profiles with respect to different polarization of the Nikon 40$\times$II Nomarski prism. (a) the superposition $I^+(x,y)+I^-(x,y)$ with polarization $(\hat{P}^+ + \hat{P}^-)/\sqrt{2}$. (b) $I^+(x,y)$ with polarization $\hat{P}^+$. (c) $I^-(x,y)$ with polarization $\hat{P}^-$. $(x^+, y^+)$ and $(x^-, y^-)$ are the central positions determined by equation \eqref{eq:GaussianFunction}. $\Delta_y=|y^+-y^-|$ denotes the spatial separation in the y direction.}\label{fig:LocalizationMethod}
\end{figure*}

\section{Results}
We demonstrate our method using the Nikon 40$\times$II Nomarski prism. Figure \ref{fig:LocalizationMethod} shows the images of intensity profiles of the refracted light beams on the Fourier plane after passing through the prism. When both $\hat{P}^+$ and $\hat{P}^-$ are present, the intensity profile is the superposition of two Gaussian beams as shown in figure \ref{fig:LocalizationMethod} (a). In this case, the spatial separation between them is too small to be resolved. By the data localization, the Gaussian beams with different polarization are acquired independently at $\omega_1$ and $\omega_2$ in figure \ref{fig:LocalizationMethod} (b) and (c). The highly overlapped light beams are completely isolated in the $\vec{k}-\hat{P}$ joint space.

Following the localization method described in section 2, we obtain the shear displacements in $x$ and $y$ directions $\Delta_x = 0.670 \pm 0.048$ $\mu$m, $\Delta_y = 8.815 \pm 0.049$ $\mu$m, and the total shear displacement $\Delta =\sqrt{\Delta_x^2+\Delta_y^2}= 8.841\pm0.049$ $\mu$m, where the uncertainties come from the fitting standard errors of the functions $g^\pm$ in equation (\ref{eq:GaussianFunction}). In both cases of $I^\pm$, the width of the fitted Gaussian function at $1/e^{2}$ gives the spot sizes $d$ around 40 $\mu$m. By comparison, the shear displacement determined by our approach is much smaller than the spot size ($\Delta \ll d$). This suggests that we are able to resolve the spatial separation of the light beams having nearly 80\% overlap in both spatial space and $\vec{k}$ space. The optical beam shear angle $\varepsilon$ is determined by the equation (\ref{eq:ShearAngle}), i.e. $\varepsilon = 58.94 \pm 0.33$ $\mu$rad, which is significantly smaller than the diffraction angle of our incident laser beam $4\lambda/(\pi D)=231$ $\mu$rad and far beyond the diffraction limit. 

\begin{figure}
\centering
\includegraphics[width=3.2in]{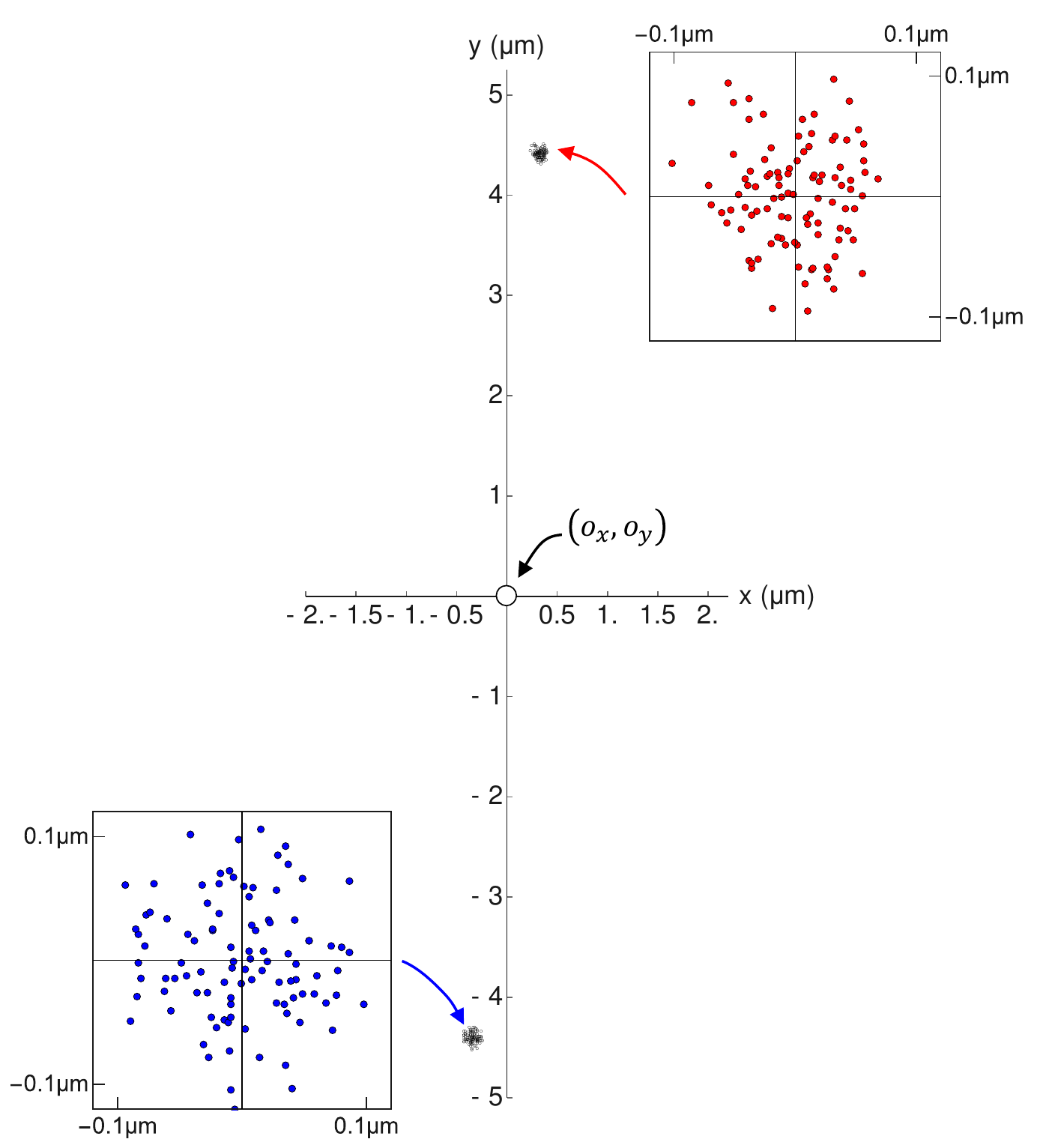}
\caption{Statistical distribution of the central positions $(x^+_i, y^+_i)$ and $(x^-_i, y^-_i)$ obtained from localization analysis.} \label{fig:LocalizationError}
\end{figure}

\section{Statistical Error Analysis}
To further validate the measurement and underlie the localization precision, we conduct the statistical error analysis with 100 frames of image at each of the polarization states. That is $I^{\pm}_i(x,y)$ for $i=1,...,100$.  For each of the images, we use the aforementioned localization analysis to determine the central positions and the corresponding optical shear displacements. Figure \ref{fig:LocalizationError} shows the scatter of all central positions with respect to $(o_x, o_y) = (\frac{\bar x^+ + \bar x^-}{2}, \frac{\bar y^+ + \bar y^-}{2})$, where $(\bar x^+, \bar y^+)$ and $(\bar x^-, \bar y^-)$ are the mean position vectors among all measured ones for $\hat P^+$ and $\hat P^-$ respectively. The insets of figure \ref{fig:LocalizationError} shows the distribution of the data measured in each of the independent polarization states. Statistically, the standard deviations of the position vectors are (46 nm, 48 nm)  and (36 nm, 42 nm) for the polarization states $\hat{P}^+$ and $\hat{P}^-$ respectively. Increasing the number of independent measurements with the application of the localization analysis does not further increase the scattering of the data. This suggests that the error raised by our experiment for the determination of the position vectors is in nanoscale. Compared with the shear displacement, i.e. $\Delta=$8.841 $\mu$m and the diffraction-limited spot size $d=$40 $\mu$m, this localization error ($\delta$) is far beyond the diffraction limit, \textit{i.e.,} $\delta\ll\Delta\ll d$. 

\begin{figure}[ht]
\centering
\includegraphics[width=3in]{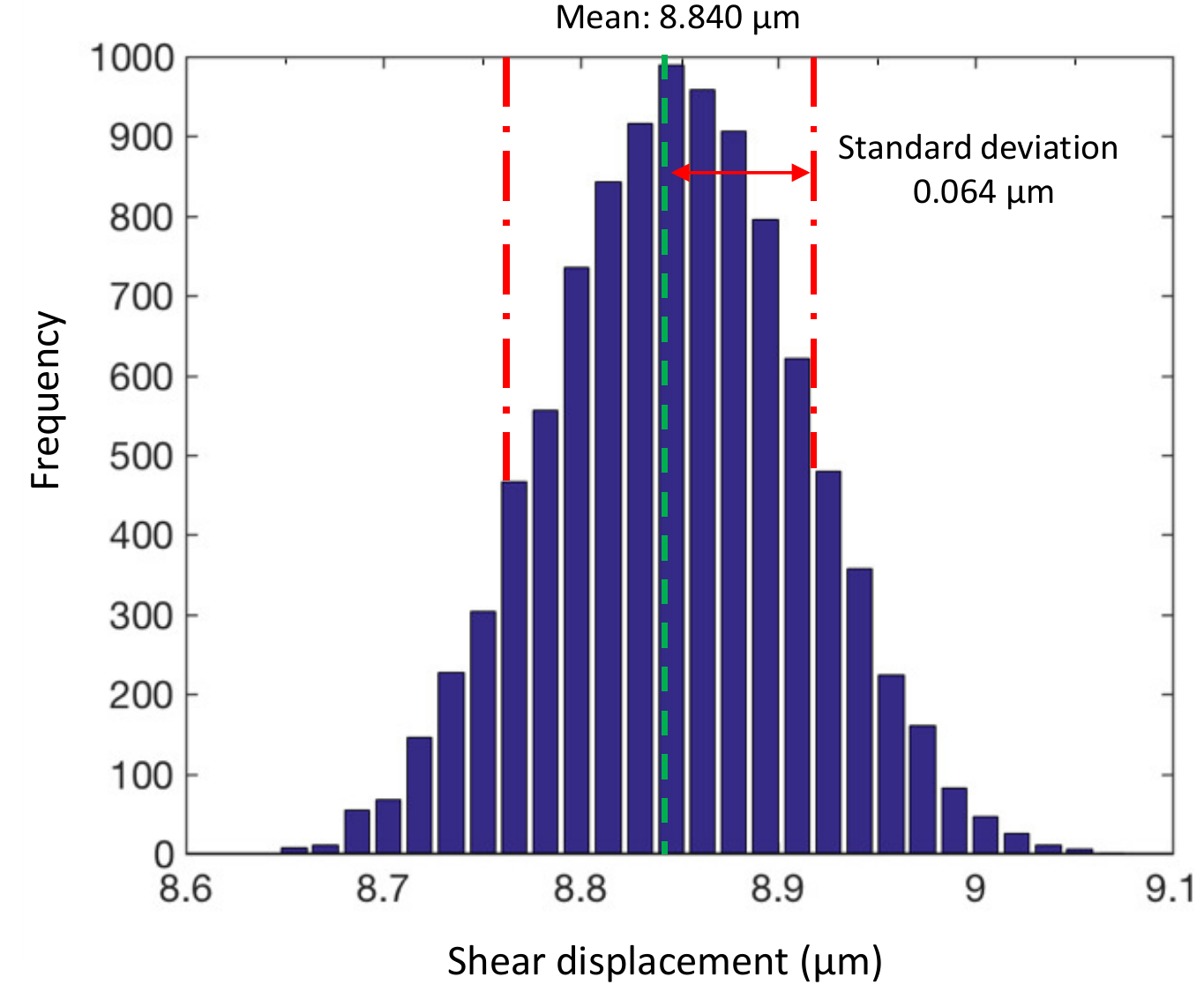}
\caption{Histogram of the shear displacement between the center positions of the two spots. The average separation distance is 8.840 $\mu$m, and the standard deviation is 0.064 $\mu$m.} \label{fig:SeparationHistogram}
\end{figure}

Bringing the experimental errors to the shear displacement, Figure \ref{fig:SeparationHistogram} shows the histogram of all calculated shear displacements 
\begin{equation}\Delta_{ij}=\sqrt{(x^+_i-x^-_j)^2+(y^+_i-y^-_j)^2},
\end{equation}
for $i, j=1,...,100$. 
The expectation of the shear displacement is $\langle\Delta_{ij}\rangle=8.840$ $\mu$m with a standard deviation of 0.064 $\mu$m. Following equation (\ref{eq:ShearAngle}), the optical beam shear angle $\varepsilon$ is $\varepsilon = 58.94 \pm 0.42$ $\mu$rad. This result agrees with that obtained from only two images and the localization fitting standard error is consistent with the statistical error. It suggests that during every measurement, the accuracy to determine the shear angle is sufficiently high for just taking two frames of $I^+(x,y)$ and $I^-(x,y)$. Table \ref{table:NikonShearAngle} lists the optical beam-shear angles of a series of Nikon Nomarski prisms determined by our method using the same setup. The angular precision of them are all less than 0.5 $\mu$rad. 

\begin{table}
\caption{The optical beam shear angles of a series of prisms}\label{table:NikonShearAngle} 
\centering
\begin{tabular}{c|c}
  \hline
Nikon Nomarski prism & Shear angle $\varepsilon$  ($\mu$rad) \\ \hline
40$\times$II & $58.94\pm0.42$ \\ \hline
60$\times$IV & $83.77\pm0.47$ \\ \hline
100$\times$I & $134.04\pm0.39$ \\ \hline
  \end{tabular}
\end{table}

\section{Theoretical precision}
The primary error of our method for characterizing the optical beam-shear angle of polarizing prisms is related to the power stability of the laser source. Unstable laser power results in scattering of central positions when performing the repeated localization measurements. This has been studied in the statistical analysis. In principle, if we only consider the effects of photon number noise and the finite pixel size, the precision of our method can be estimated as \cite{ThompsonBJ2002}.
\begin{eqnarray}
\delta\approx\sqrt{\frac{\sigma^2+(a^2/12)}{N}+\frac{4\sqrt{\pi \sigma^3 b^2}}{a N^2}},
\label{eq:precisiontheory}
\end{eqnarray}
where $\sigma=d/4$ is the standard deviation of the single Gaussian spot of intensity distribution, $a$ is the camera pixel size, $N$ is the number of photons collected, and $b$ is the background noise. In our experiment, $d=40.8$ $\mu$m, $a=2.4$ $\mu$m, $N=308600$, and $b\simeq0$ by which the theoretical localization error is $\delta\approx0.018$ $\mu$m and shear angle error is $\delta_{\varepsilon}=\sqrt{2}\delta/f\approx0.17$ $\mu$rad. Admittedly, the error in our experiment is greater than this theoretical value by a factor of 3. This is attributed to the unavoidable perturbations from the environment, the alignment of the half-wave plate with respect to the prism, the mechanical vibration, and the temperature fluctuation in the lab. In our experiment, we took 100 frames for $\hat{P}^{+}$ firstly and then next 100 frames for $\hat{P}^{-}$.  The effect of the perturbations are added up during this sequential data acquisition process, which can be reduced if we can replace the half-wave plate with a fast polarization switch (such as an electrically controlled liquid crystal polarization rotator) and take the two frames for $I^+(x,y)$ and $I^-(x,y)$ consequentially within a short time interval. In this way, although the position vectors of the centroids are scattered at the same level, the numerical error from the determination of the correlated shear displacement $\Delta_{ii}$ would be reduced. In addition, using the camera with a higher saturation level for receiving more photons (i.e. the value $N$ in Eq. (\ref{eq:precisiontheory})) in each frame, a nanometer localization accuracy would be achieved \cite{GellesNature1988}, which may lead to a measurement precision of $0.01$ $\mu$rad.

\section{Conclusion}
In summary, we proposed a method that precisely characterizes the optical beam-shear angle of polarizing prisms with the application of the localization analysis. Using a Fourier lens, the two orthogonally polarized beams after the prism are transformed into the $\vec{k}-\hat{P}$ joint space, where the central positions of each of the polarization modes can be determined independently. Since the polarization states are distinguishable, it is possible to directly measure the spatial separations of two light beams with almost 80\% overlapping. We demonstrated our method in a series of Nikon Nomarski prisms and achieved the angular precision within 0.5 $\mu$rad. We also discussed the theoretical precision limit, which suggests that a much higher precision can be achieved in an improved system. The result ($\varepsilon = 58.94 \pm 0.33$ $\mu$rad) obtained from the two images data in figure \ref{fig:LocalizationMethod}(b) and (c)  agrees well with that ($\varepsilon = 58.94 \pm 0.42$ $\mu$rad) from the statistical analysis shown in Figs. \ref{fig:LocalizationError} and \ref{fig:SeparationHistogram}. This validates that only two images [as figure \ref{fig:LocalizationMethod}(b) and (c)] are sufficient to determine the optical beam-shear angle with a high accuracy. The design of optical setup is a stand-alone experiment that can be applied to many other polarization-based birefringent devices.

\section*{Funding}
Hong Kong University of Science and Technology (HKUST) R and D Corporation Ltd (Project Code 16171510); Hong Kong Research Grants Council (Project Nos ECS26200136 and GRF16207017); Hong Kong University Grants Committee (IGN16EG04).

\section*{Acknowledgments}
The work was supported by Light Innovation Technology Ltd through a research contract with the Hong Kong University of Science and Technology (HKUST) R and D Corporation Ltd (Project Code 16171510). X.C. acknowledges the financial support from the HK Research Grants Council through ECS Grant 26200136, the UGC Grant IGN16EG04 and the GRF Grant 16207017. H. C. C., supervised by S. D. and X. C. thanks the Undergraduate Research Opportunities Program at HKUST.

\bibliographystyle{model1-num-names}
\bibliography{prism_ref.bib}







\end{document}